\documentclass{article}

\date{} 

\usepackage{enumitem}

\usepackage{url}

\usepackage{hyperref} 

\usepackage{spconf,amsmath,graphicx}
\usepackage{amssymb}
\usepackage{algorithm,algorithmic}
\usepackage{url}
\makeatletter
\newcommand{\ALOOP}[1]{\ALC@it\algorithmicloop\ #1%
  \begin{ALC@loop}}
\newcommand{\ENDALOOP}{\end{ALC@loop}\ALC@it\algorithmicendloop}

\makeatother

\usepackage{setspace}
\usepackage[skip=2pt, font=normalsize]{caption}
\newcommand*{\Scale}[2][4]{\scalebox{#1}{$#2$}} 

\title{Tensor-based subspace learning for tracking salt-dome boundaries}
\name{Zhen Wang, Zhiling Long, and Ghassan AlRegib}
\address{Center for Energy and Geo Processing (CeGP) at Georgia Tech and KFUPM\\
School of Electrical and Computer Engineering\\
Georgia Institute of Technology, Atlanta, GA 30332-0250, USA\\
\{zwang313, zhiling.long, alregib\}@gatech.edu}
\begin{document}

\onecolumn 

\begin{description}[labelindent=1cm,leftmargin=4cm,style=multiline]

\item[\textbf{Citation}]{Z. Wang, Z. Long and G. AlRegib, ``Tensor-based subspace learning for tracking salt-dome boundaries,'' Proceedings of IEEE ICIP 2015, Québec City, Canada, Sep. 2015.}
\\
\item[\textbf{DOI}]{\url{https://doi.org/10.1109/ICIP.2015.7351083}}
\\
\item[\textbf{Review}]{Date of publication: 10 December 2015}
\\
\item[\textbf{Data and Codes}]{\url{https://ghassanalregib.com/}}
\\
\item[\textbf{Bib}] {@INPROCEEDINGS\{7351083, \\ 
author=\{Z. Wang and Z. Long and G. AlRegib\}, \\ 
booktitle=\{2015 IEEE International Conference on Image Processing (ICIP)\}, \\ 
title=\{Tensor-based subspace learning for tracking salt-dome boundaries\}, \\ 
year=\{2015\}, \\ 
pages=\{1663-1667\}, \\ 
doi=\{10.1109/ICIP.2015.7351083\}, \\ 
month=\{Sep.\}\} \\
} 


\item[\textbf{Copyright}]{\textcopyright 2015 IEEE. Personal use of this material is permitted. Permission from IEEE must be obtained for all other uses, in any current or future media, including reprinting/republishing this material for advertising or promotional purposes, creating new collective works, for resale or redistribution to servers or lists, or reuse of any copyrighted component of this work in other works.}
\\
\item[\textbf{Contact}]{\href{mailto:zhiling.long@gatech.edu}{zhiling.long@gatech.edu}  OR \href{mailto:alregib@gatech.edu}{alregib@gatech.edu}\\ \url{https://ghassanalregib.com/} \\ }
\end{description}

\thispagestyle{empty}
\newpage
\clearpage
\setcounter{page}{1}

\twocolumn

%

%
\maketitle
\begin{abstract}
The exploration of petroleum reservoirs has a close relationship with the identification of salt domes. To efficiently interpret salt-dome structures, in this paper, we propose a method that tracks salt-dome boundaries through seismic volumes using a tensor-based subspace learning algorithm. We build texture tensors by classifying image patches acquired along the boundary regions of seismic sections and contrast maps. With features extracted from the subspaces of texture tensors, we can identify tracked points in neighboring sections and label salt-dome boundaries by optimally connecting these points. Experimental results show that the proposed method outperforms the state-of-the-art salt-dome detection method by employing texture information and tensor-based analysis.

\end{abstract}
\begin{keywords}
salt-dome tracking, contrast attribute, texture tensors, and subspace learning
\end{keywords}
\vspace{-0.1in}
\section{Introduction}
\label{sec:intro}

The deposition of salt in marine basins commonly intrudes into surrounding rock strata and forms an important geological structure, salt domes. Because of the impermeability of salt, salt domes may seal porous reservoir rocks and lead to the formation of petroleum reservoirs. To estimate the possible positions of reservoir regions, experienced interpreters need to accurately delineate the boundaries of salt domes in collected seismic data. With the dramatically increasing amount of acquired seismic data, however, manual interpretation is becoming time consuming and labor intensive.

To speed up interpretation, in recent years, researchers have proposed various computer-aided methods that detect salt-dome boundaries using graph theory and image-processing techniques. Lomask et al.~\cite{lomask2004image} defined seismic sections as weighted undirected graphs and applied the normalized cut image segmentation (NCIS) to globally optimize the delineation of salt-dome boundaries. The method in \cite{lomask2007application}, an extension of~\cite{lomask2004image}, employed local dips in the weight matrix and utilized bound constraints to remove boundary artifacts. Similarly, Harpert et al.~\cite{halpert2009seismic} introduced the modified NCIS by combining multiple seismic attributes with adaptive weights. Although these NCIS-based methods can be implemented in parallel, their high computational cost limits their future application on high-resolution or three-dimensional (3D) seismic data. To improve the efficiency of global segmentation, Harpert et al.~\cite{halpert2010speeding} proposed to detect the boundaries of salt domes using the pairwise region comparison based on the minimum spanning tree, which reduces the algorithm complexity from $\mathcal{O}\left(n^2\right)$ to $\mathcal{O}\left(n\log n\right)$. In~\cite{aqrawi2011detecting}, Aqrawi et al. applied a dip-guided 3D Sobel filter to detect salt-dome boundaries. Recently, Berthelot et al.~\cite{berthelot2013texture} have introduced to segment salt domes by testing the combination of multiple seismic attributes--texture attributes, frequency-based attributes, and dip attributes--in the supervised Bayesian classification model. In~\cite{hegazy2014texture}, to characterize texture differences between salt domes and surrounding geological structures, Hegazy and AlRegib proposed the directionality attribute using the moment of inertia tensors of gradient components

In this paper, we propose a new algorithm for tracking the boundaries of salt domes through 3D seismic volumes. We first derive the contrast map from an initial image, the salt-dome boundary of which has been accurately labeled by experienced interpreters or practical detection methods. On the basis of the initial image and its corresponding contrast map, we build third-order texture tensors along the labeled boundary. Then, by employing features extracted from the subspace of texture tensors, we identify the positions of tracked boundary points in neighboring images. Finally, the optimal connection of tracked points synthesizes tracked boundaries.
To maintain consistency with the terminology in video coding, we rename initial and neighboring images as reference and predicted sections, respectively, in the rest of this paper.

\vspace{-0.1in}
\section{The Proposed Method}
\label{sec:method}
\begin{figure}[t]
\begin{minipage}[b]{1.0\linewidth}
  \centering
  \centerline{\includegraphics[width=8.5cm]{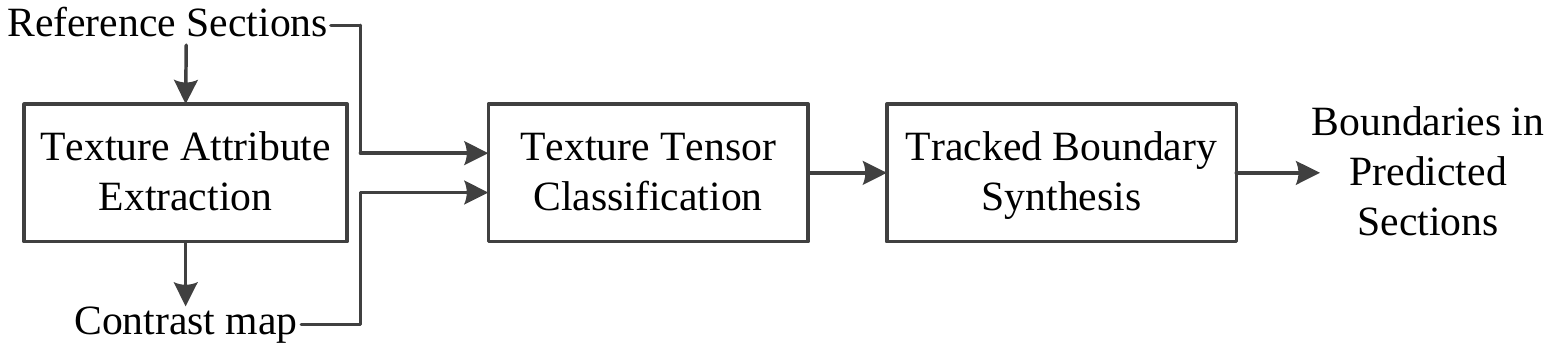}}
\end{minipage}
\caption{Block diagram of the proposed method.}
\label{fig:diagram}
\end{figure}

\vspace{-0.1in}
The block diagram of the proposed method is shown in Fig.~\ref{fig:diagram}. In the following subsections, we explain each step of the proposed pipeline in detail.
\subsection{Texture Attribute Extraction}
\label{ssec:attribute}

Since rock strata surrounding salt domes are commonly formed by sedimentary rocks such as shale and limestone, the various properties of salt and sedimentary rocks result in their distinctive appearances in seismic sections. Fig.~\ref{fig:fig1}(a) illustrates a reference section obtained from the North Sea dataset~\cite{f3opendtect} with inline number 399, in which a blue curve indicates the boundary of a salt dome. By observing textures along the labeled boundary, we notice that in contrast to the homogeneous textures of the salt body, neighboring rock strata have varying, more complicated textures. Such a prominent texture contrast represents an important feature of salt-dome boundaries.

To accurately characterize boundary textures, we employ a gray level co-occurrence matrix (GLCM)~\cite{haralick1973textural} that describes the distribution of co-occurring grayscale values at a given offset over an image. For each point in the reference section, we calculate its corresponding GLCM within an $(2R_d+1)\times(2R_d+1)$ analysis window as $\Scale[0.87]{G_{\left[\Delta x, \Delta y\right]}[i,j] = \mbox{Pr}\left(S[x,y] = i, S\left[x+\Delta x, y+\Delta \right]=j\right)}$,
where $x$ and $y$ correspond to the indices along the crossline and depth directions, respectively, and $S[x,y]$ represents the intensity of seismic signals. $\mbox{Pr}(\cdot)$ denotes the probability of an event. In addition, $i$ and $j$ correspond to grayscales under specified quantization, and $\left[\Delta x, \Delta y\right]$ defines the predetermined offset, which is limited to four directions, $\left\{0^\circ, 45^\circ, 90^\circ, 135^\circ\right\}$ and has a pixel distance within the range of $1$ to $R_d$. By selecting various directions and pixel distances, we can derive $N_g$ GLCMs for each point that contain multi-directional and multi-scale texture information. From obtained GLCMs, we extract the contrast attribute that describes the variation of textures as follows:
\vspace{-0.07in}
\begin{equation}
\label{eq:glcmCont}
\Scale[0.93]{
\overline{C}[x,y] = \sum\limits_{\left[\Delta x, \Delta y\right]}\left(\frac{1}{N_g}\sum\limits_{i}\sum\limits_{j}(i-j)^2G_{\left[\Delta x, \Delta y \right]}[i,j]\right).}
\end{equation}

\vspace{-0.07in}
\noindent Fig.~\ref{fig:fig1}(b) illustrates the contrast map of Inline \#399. Blue areas with contrast values close to zero represent homogeneous textures. In contrast, green and red stripes with contrast values close to one indicate great texture variations around the salt-dome boundary.

\vspace{-0.15in}
\subsection{Texture Tensor Classification}
\label{ssec:tensor}

\vspace{-0.05in}
\subsubsection{Tensors and Multi-linear Analysis}
\label{sssec:tensorIntro}

\vspace{-0.05in}
In the field of multi-linear algebra, tensors are commonly used to describe high-dimensional ($N\geq 3$) data.
For an $N$-th order tensor $\mathcal{A}\in\mathbb{R}^{I_1\times I_2\times\cdots\times I_N}$, each order represents a mode of $\mathcal{A}$. We can unfold $\mathcal{A}$ along the $n$-mode and obtain matrix $\mathbf{A}^{(n)}\in\mathbb{R}^{I_n\times(I_1\times\cdots\times I_{n-1}\times I_{n+1}\cdots\times I_N)}$~\cite{de2000best}. 
The $n$-mode product of tensor $\mathcal{A}$ by matrix $\mathbf{U}\in\mathbb{R}^{J_n\times I_n}$, denoted $\mathcal{A}\times_n\mathbf{U}$, is new tensor $\mathcal{B}\in\mathbb{R}^{I_1\times\cdots\times I_{n-1}\times J_n\times I_{n+1}\cdots\times I_N}$, the entries of which are calculated as
$\mathcal{B}(i_1\cdots i_{n-1}j_ni_{n+1}\cdots i_N)=\sum_{i_n}\mathcal{A}(i_1\cdots i_N)\cdot \mathbf{U}(j_n,i_n)$.
Tensors can be decomposed as $\mathcal{A}=\mathcal{S}\times_1\mathbf{U}^{(1)}\times_2\mathbf{U}^{(2)}\cdots\times_N\mathbf{U}^{(N)}$, in which $\mathbf{U}^{(n)}$, $n=1,2,\cdots,N$, represents an $I_n\times I_n$ orthogonal matrix and its column vectors span the column space of $n$-mode unfolding matrix $\mathbf{A}^{(n)}$. Since image and video tensors commonly have high dimensions along each mode, we apply multi-linear principal component analysis (MPCA)~\cite{lu2008mpca} to extract a tensor subspace with lower dimensions that captures the most variation of input data. The projection operation can be expressed as $\mathcal{C}=\mathcal{A}\times_1\widetilde{\mathbf{U}}^{(1)^T}\times_2\widetilde{\mathbf{U}}^{(2)^T}\cdots\times_N\widetilde{\mathbf{U}}^{(N)^T}\in\mathbb{R}^{P_1\times P_2\times\cdots\times P_n}$. $\widetilde{\mathbf{U}}^{(n)}$, $n=1,2,\cdots,N$, represent $I_n\times P_n$ matrices, the column vectors of which are orthonormal and composed of eigenvectors corresponding to the largest $P_n$ eigenvalues of $\mathbf{A}^{(n)}\cdot\mathbf{A}^{(n)^T}$.\\




\vspace{-0.3in}
\subsubsection{Adaptive Classification of Texture Tensors}
\label{sssec:ini}
\begin{figure}[t]
\begin{minipage}[b]{0.48\linewidth}
  \centering
  \centerline{\includegraphics[width=4.2cm]{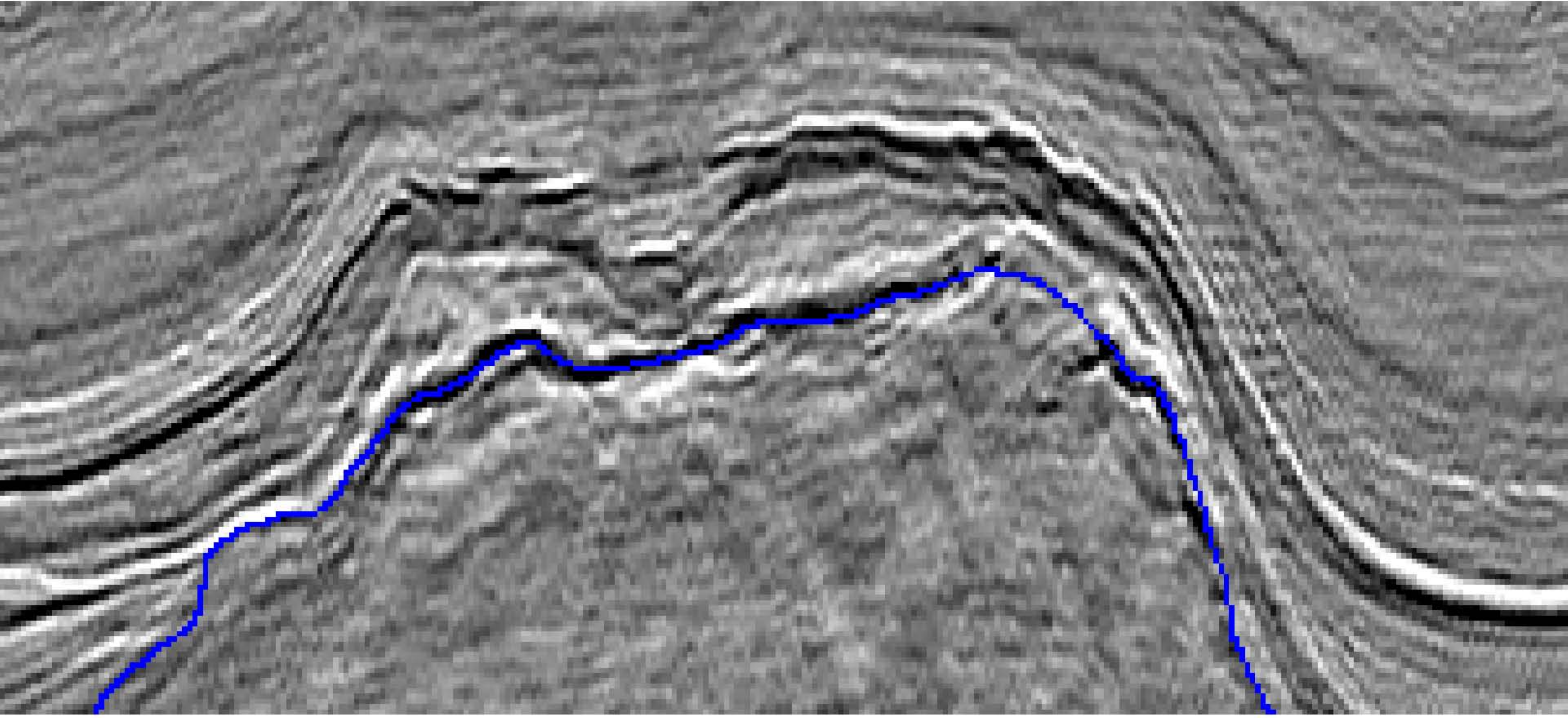}}
  \centerline{(a) Inline \#399}\medskip
\end{minipage}
\hfill
\begin{minipage}[b]{0.48\linewidth}
  \centering
  \centerline{\includegraphics[width=4.6cm]{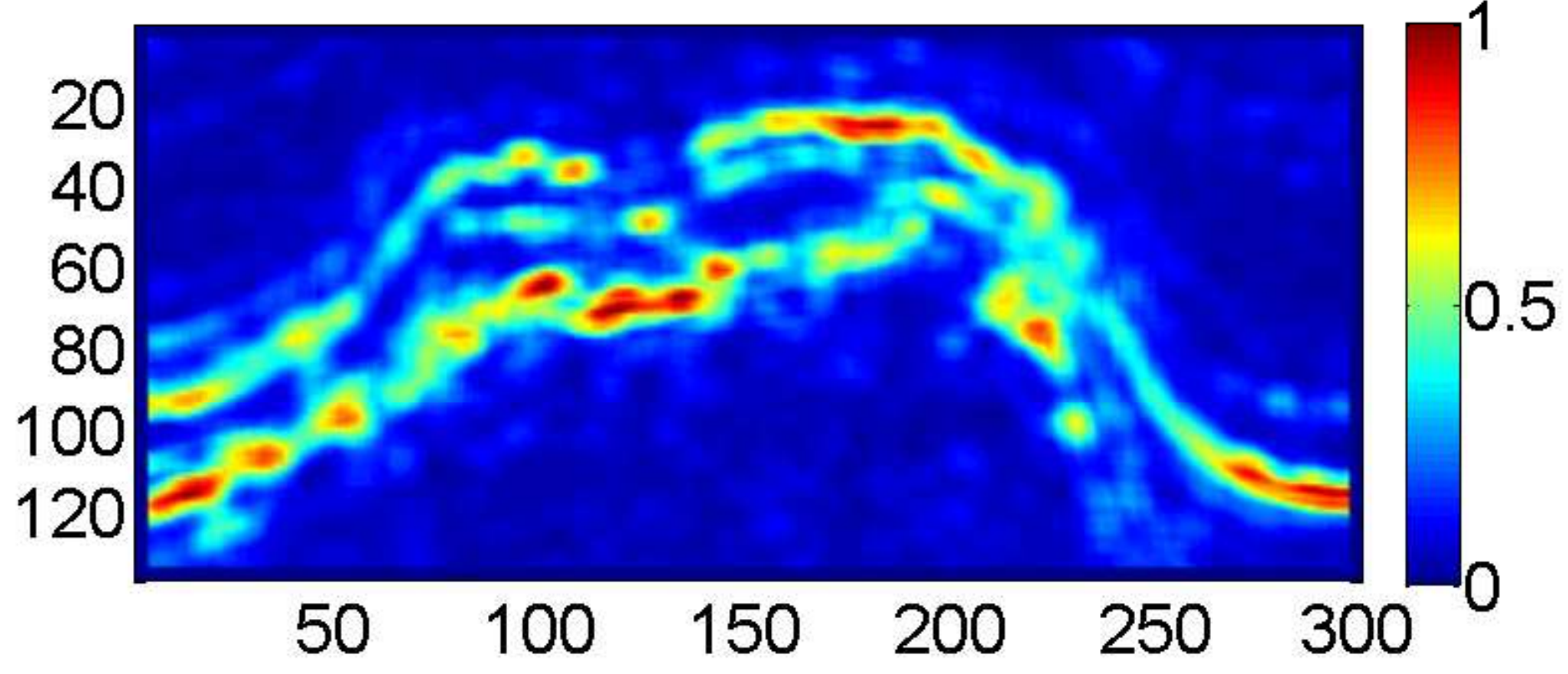}}
  \centerline{(b) GLCM contrast map}\medskip
\end{minipage}
\caption{An example of a seismic section, Inline \#399, and its corresponding GLCM contrast map.}
\label{fig:fig1}
\end{figure}
To fully utilize texture information along salt-dome boundaries, we extract pairs of patches centered at boundary points, $p_i$, $i=1,2,\cdots,N_b$, from the reference section and its corresponding contrast map. $N_b$ defines the number of all boundary points. Since matrices are special third-order tensors with $I_3=1$, patch pairs, denoted $\left\{\mathcal{S}_{p_i}, \mathcal{C}_{p_i}\right\}\subset\mathbb{R}^{I_1\times I_2\times1}$, contain two types of third-order texture tensors. $I_1$ and $I_2$ represent the dimensions of patches along the crossline and depth directions, respectively, which are empirically determined based on the complexity of boundary textures. Since the formation of salt domes lasts hundreds of years, local areas commonly have strong uniformity, which is an important constraint to classifying textures along the boundaries of salt domes.

\begin{figure}[t]
\begin{minipage}[b]{0.25\linewidth}
  \centering
  \centerline{\includegraphics[height=2.6cm]{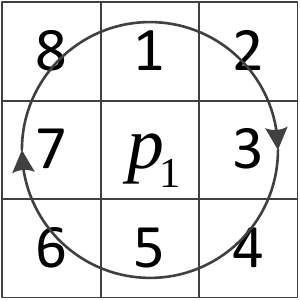}}
  \centerline{(a)}\medskip
\end{minipage}
\hfill
\begin{minipage}[b]{0.75\linewidth}
  \centering
  \centerline{\includegraphics[height=2.7cm]{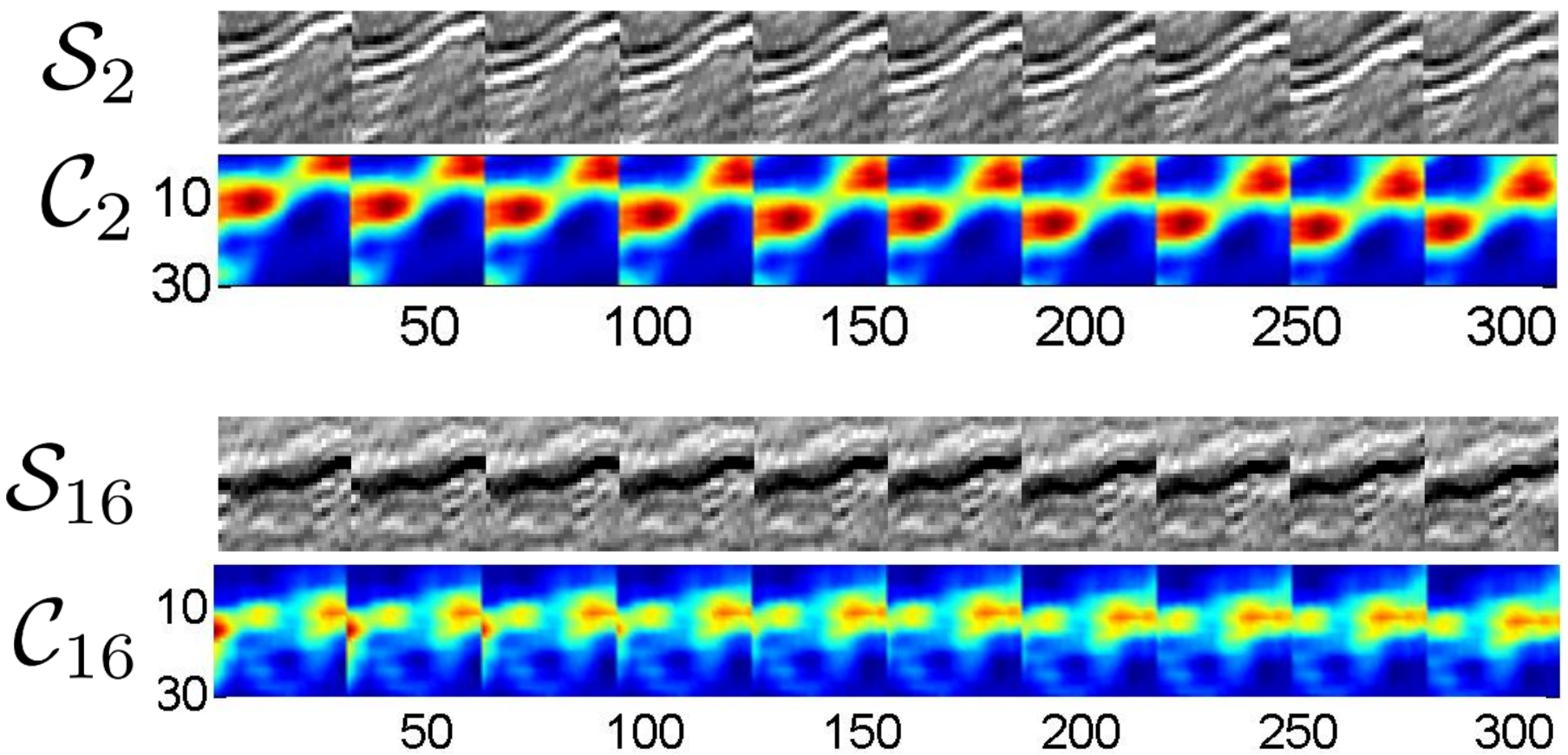}}
  \centerline{(b)}\medskip
\end{minipage}
\caption{(a) The priority of neighboring points, and (b) patches in two tensor pairs, $\{\mathcal{S}_2,\mathcal{C}_2\}$, and $\{\mathcal{S}_{16},\mathcal{C}_{16}\}$.}
\label{fig:fig4}
\end{figure}

In this section, we attempt to build pairs of tensors, $\{\mathcal{S}_k, \mathcal{C}_k\}$, $k=1,2,\cdots,N_t$, constructed by the grouped patches of boundary points containing similar textures. The pseudocode of the proposed classification method is listed in Algorithm~\ref{algo:algo1}.
\begin{algorithm}
  \caption{The Classification of Texture Tensors}\label{algo:algo1}
  \begin{algorithmic}[1]
    \REQUIRE a set of patch pairs $\Scale[0.96]{\{(\mathcal{S}_{p_i},\mathcal{C}_{p_i}),i=1,2,\cdots,N_b\}}$
    \ENSURE texture tensor pairs $\Scale[0.96]{\{(\mathcal{S}_k,\mathcal{C}_k),k=1,2,\cdots,N_t\}}$
    \STATE \textbf{Initialization:} $\Scale[0.96]{\mathcal{S}_1\gets \mathcal{S}_{p_1}}$, $\Scale[0.96]{\mathcal{C}_1\gets \mathcal{C}_{p_1}}$, and $k=1$
        \FOR {$i\gets 2\mbox{ to }N_b$} 
        \STATE
        $\Scale[0.6]{\bullet}$ $\widetilde{\mathcal{S}}_{k}\gets (\mathcal{S}_k|\mathcal{S}_{p_i})\mbox{, and }\widetilde{\mathcal{C}}_{k}\gets (\mathcal{C}_k|\mathcal{C}_{p_i})$\\
        \STATE $\Scale[0.6]{\bullet}$ Calculate $\Scale[0.96]{\left\{\widetilde{\mathbf{U}}_{\widetilde{\mathcal{M}}_k}^{(n)},\ n=1,2,3,\  \mathcal{M}=\{\mathcal{S},\mathcal{C}\}\right\}}$\\
        \STATE $\Scale[0.6]{\bullet}$ Calculate reconstruction errors:\\
        \STATE $\Scale[0.80]{
        \begin{aligned}
        e=&\sum\limits_{n=\{1,2\},\mathcal{M}=\{\mathcal{S},\mathcal{C}\}}\left\|\mathcal{M}_{p_i}-\mathcal{M}_{p_i}\times_n\left(\widetilde{\mathbf{U}}^{(n)}_{\widetilde{\mathcal{M}}_k}\cdot\widetilde{\mathbf{U}}^{(n)^{T}}_{\widetilde{\mathcal{M}}_k}\right)\right\|_{F}^2+\\
        &\sum\limits_{\mathcal{M}=\{\mathcal{S},\mathcal{C}\}}\left\|\mathbf{M}_{p_i}^{(3)}-\mathbf{M}_{p_i}^{(3)}\cdot\widetilde{\mathbf{U}}_{\widetilde{\mathcal{M}}_k}^{(3)}\cdot\widetilde{\mathbf{U}}_{\widetilde{\mathcal{M}}_k}^{(3)^T}\right\|_F^2   \end{aligned}}$

        \IF {$e\leq T_e$}
            \STATE $\mathcal{S}_k\gets\widetilde{\mathcal{S}}_k\mbox{, and }\mathcal{C}_k\gets\widetilde{\mathcal{S}}_k$
        \ELSE
            \STATE $k\gets k+1$
            \STATE $\Scale[0.96]{\mathcal{S}_k\gets \mathcal{S}_{p_i}}$, $\Scale[0.96]{\mathcal{C}_k\gets \mathcal{C}_{p_i}}$\\
        \ENDIF
    \ENDFOR
  \end{algorithmic}
\end{algorithm}
We first initialize $\{\mathcal{S}_1,\mathcal{C}_1\}$ using patches $\{\mathcal{S}_{p_1},\mathcal{C}_{p_1}\}$. Since in seismic sections the boundaries of salt domes commonly appear in roughly semi-circular shapes, to transverse all boundary points clockwise, we define $p_1$ as the first point in the bottom-left corner satisfying the dimension constraint of patches. To identify the next boundary point $p_2$, we search clockwise in the $3\times 3$ neighborhood of $p_1$. Fig.~\ref{fig:fig4}(a) illustrates the priority of neighboring points, in which a smaller number represents higher priority. After determining the position of $p_2$, we append the corresponding patches $\{\mathcal{S}_{p_2},\mathcal{C}_{p_2}\}$ to the current tensor pair $\{\mathcal{S}_1,\mathcal{C}_1\}$ along the $z$ direction and denote updated tensors as $\widetilde{\mathcal{S}}_1$ and $\widetilde{\mathcal{C}}_1$, respectively. By applying the MPCA on tensors $\{\widetilde{\mathcal{S}}_1,\widetilde{\mathcal{C}}_1\}$, we obtain the projection matrices of all modes, denoted $\widetilde{\mathbf{U}}_{\widetilde{\mathcal{M}}_1}^{(n)}\in\mathbb{R}^{I_n\times P_n}$, $(n=1,2)$, $\widetilde{\mathbf{U}}_{\widetilde{\mathcal{M}}_1}^{(3)}\in\mathbb{R}^{(I_1\times I_2)\times P_n}$,  $\mathcal{M}=\{\mathcal{S},\mathcal{C}\}$, in which $[P_1,P_2,P_3]$ are determined empirically. For 3-mode unfolding matrices, to utilize the high computational efficiency of the sequential Karhunen-Loeve (SKL) algorithm~\cite{levey2000sequential,hu2011incremental}, we extract the projection matrices of row spaces. On the basis of reconstruction error $e$, we can evaluate the similarity between $\{\mathcal{S}_{p_2},\mathcal{C}_{p_2}\}$ and $\{\mathcal{S}_1, \mathcal{C}_1\}$. If $e$ is less than threshold $T_e$, high similarity leads to the extension of current tensors with $\{\mathcal{S}_{p_2},\mathcal{C}_{p_2}\}$ appended. Otherwise, $\{\mathcal{S}_{p_2},\mathcal{C}_{p_2}\}$ initializes the next new tensor pair $\{\mathcal{S}_{2}, \mathcal{C}_{2}\}$. By repeating the steps above, we obtain classified texture tensors. Fig.~\ref{fig:fig4}(b) illustrates the patches of two tensor pairs, the strong correlation among which can be better captured by tensor-based analysis.
\vspace{-0.2in}
\subsection{Tracked Boundary Synthesis}
\label{ssec:trackBoundary}

\vspace{-0.05in}
On the basis of projection matrices extracted from classified tensors, we can localize tracked boundary points in predicted sections. We first project the labeled boundary onto the target predicted section and keep the coordinates of the boundary points unchanged. Then, at each projected boundary point $p$, we search for the tracked point along its normal direction within a $(2R_s+1)$ window, in which $R_s$ represents the inline number difference between the reference section and the predicted section. Each candidate point corresponds to a pair of texture patches, denoted $\{\mathcal{S}_{p,j}, \mathcal{C}_{p,j}\}\subset\mathbb{R}^{I_1\times I_2\times 1}$, $j=1,2,\cdots,(2R_s+1)$, extracted from the predicted section and the corresponding contrast map. We assume that the texture patches of current projected point $p$ belong to tensor pair $\{\mathcal{S},\mathcal{C}\}$. The localization of tracked point $p^{\ast}$ is explained in the following pseudocode:
\vspace{-0.1in}
\begin{algorithm}
  \caption{The Localization of Tracked Points}\label{algo:algo2}
  \begin{algorithmic}[1]
    \REQUIRE a set of patch pairs $\Scale[0.85]{\{(\mathcal{S}_{p,j},\mathcal{C}_{p,j}),j=1,2,\cdots,(2R_s+1)\}}$
    \ENSURE tracked point $p^{\ast}$ and updated tensors $\{\mathcal{S}, \mathcal{C}\}$
    \STATE \textbf{Initialization:} $e_{min}=+\infty$
        \FOR {$j\gets 1\mbox{ to }(2R_s+1)$} 
        \STATE
        $\Scale[0.6]{\bullet}$ $\widetilde{\mathcal{S}}\gets (\mathcal{S}|\mathcal{S}_{p,j})\mbox{, and }\widetilde{\mathcal{C}}\gets (\mathcal{C}|\mathcal{C}_{p,j})$\\
        \STATE $\Scale[0.6]{\bullet}$ Calculate $\Scale[0.96]{\left\{\widetilde{\mathbf{U}}_{\widetilde{\mathcal{M}}}^{(n)},\ n=1,2,3,\  \mathcal{M}=\{\mathcal{S},\mathcal{C}\}\right\}}$\\
        \STATE $\Scale[0.6]{\bullet}$ Calculate reconstruction errors:\\
        \STATE $\Scale[0.77]{
        \begin{aligned}
        e=&\sum\limits_{n=\{1,2\},\mathcal{M}=\{\mathcal{S},\mathcal{C}\}}\lambda_{\mathcal{M}}\cdot\left\|\mathcal{M}_{p,j}-\mathcal{M}_{p,j}\times_n\left(\widetilde{\mathbf{U}}^{(n)}_{\widetilde{\mathcal{M}}}\cdot\widetilde{\mathbf{U}}^{(n)^{T}}_{\widetilde{\mathcal{M}}}\right)\right\|_{F}^2+\\
        &\sum\limits_{\mathcal{M}=\{\mathcal{S},\mathcal{C}\}}\lambda_{\mathcal{M}}\cdot\left\|\mathbf{M}_{p,j}^{(3)}-\mathbf{M}_{p,j}^{(3)}\cdot\widetilde{\mathbf{U}}_{\widetilde{\mathcal{M}}}^{(3)}\cdot\widetilde{\mathbf{U}}_{\widetilde{\mathcal{M}}}^{(3)^T}\right\|_F^2 \end{aligned}}$
        \IF {$e\leq e_{min}$}
            \STATE $e_{min}\gets e$, $\mathcal{S}_{p^{\ast}}\gets \mathcal{S}_{p,j}$, and $\mathcal{C}_{p^{\ast}}\gets \mathcal{C}_{p,j}$
        \ENDIF
    \ENDFOR
    \STATE $\mathcal{S}\gets (\mathcal{S}|\mathcal{S}_{p*})\mbox{, and }\mathcal{C}\gets (\mathcal{C}|\mathcal{C}_{p*})$\\
  \end{algorithmic}
\end{algorithm}

\begin{figure}[t]
\begin{minipage}[b]{0.48\linewidth}
  \centering
  \centerline{\includegraphics[height=2cm]{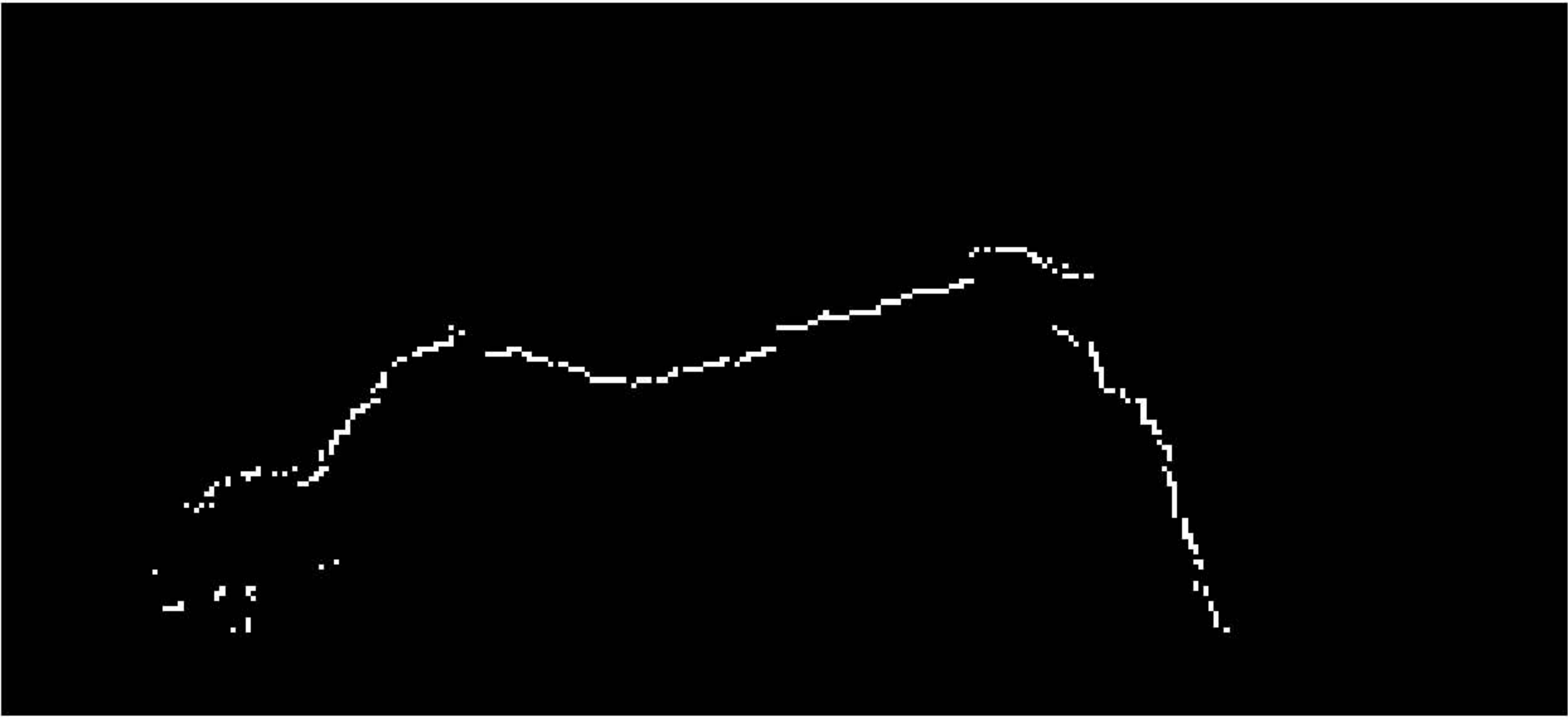}}
  \centerline{(a)}\medskip
\end{minipage}
\hfill
\begin{minipage}[b]{0.48\linewidth}
  \centering
  \centerline{\includegraphics[height=2cm]{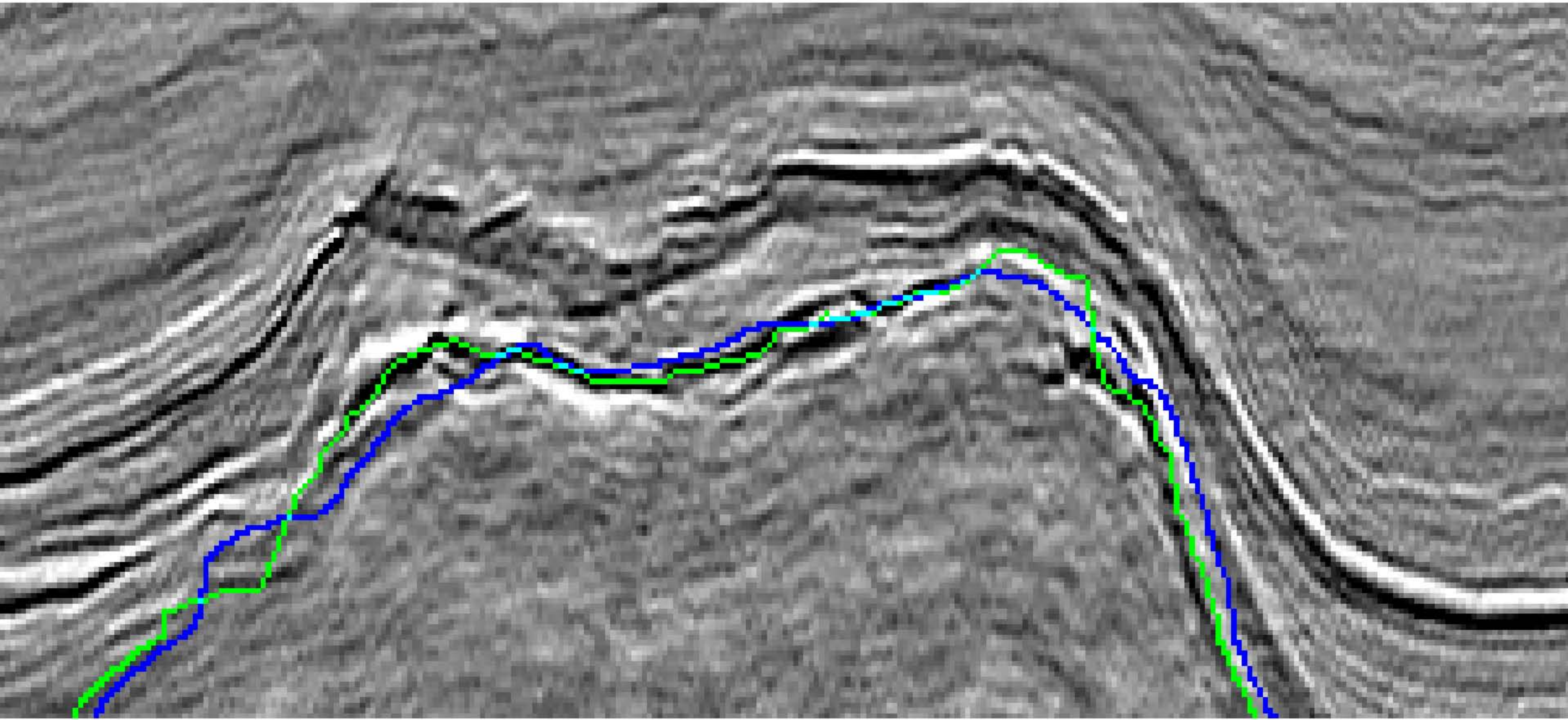}}
  \centerline{(b)}\medskip
\end{minipage}
\caption{(a) Tracked points in Inline \#409, and (b) green tracked salt-dome boundary in Inline \#409 and blue boundary projected from the reference section Inline \#399.}
\label{fig:fig5}
\end{figure}

\vspace{-0.1in}
\noindent By evaluating the similarity between current tensors $\{\mathcal{S},\mathcal{C}\}$ and the patch pairs of the candidate points, we can localize tracked point $p^{\ast}$. $\lambda_{\mathcal{S}}$ and $\lambda_{\mathcal{C}}$ represent the weights of $\mathcal{S}$ and $\mathcal{C}$ in reconstruction error $e$. Since the ranges of seismic sections and contrast maps have been normalized between $0$ and $1$, we define $\lambda_{\mathcal{S}}=1$ and $\lambda_{\mathcal{C}}=|\log(\overline{C})|$. Therefore, candidate points with higher contrast values have lower weights in $e$. Fig.~\ref{fig:fig5}(a) illustrates tracked points in Inline \#409. After removing noisy points using the $2\times 2$ median filter, we connect the remaining tracked points to synthesize the tracked boundary. In Fig.~\ref{fig:fig5}(b), green and blue curves represent the tracked and projected boundary in Inline \#409, respectively.

\vspace{-0.1in}
\section{Experimental Results}
\label{sec:results}

In this paper, we apply the proposed salt-dome tracking method on a 3D real seismic dataset acquired from the Netherlands offshore F3 block with the size of $24\times 16$ km$^{2}$ in the North Sea~\cite{f3opendtect}. To illustrate the performance of the proposed method, we focus on a local volume of the dataset containing discernible salt-dome structures. The tested volume has an inline number ranging from \#151 to \#501, a crossline number ranging from \#401 to \#701, and a time direction ranging from 1,300ms to 1,848ms with a step of 4ms. Both Figs.~\ref{fig:fig1}(a) and~\ref{fig:fig5}(b) illustrate seismic sections extracted from the local volume.

As we mentioned in previous sections, on the basis of the labeled salt-dome boundary in the reference section Inline \#399, we attempt to synthesize tracked salt-dome boundaries in neighboring predicted sections ranging from Inline \#389 to \#409. To acquire the contrast map of each seismic section, we derive the GLCMs of points from $9\times 9$ neighborhoods by selecting various directions and pixel distances. Then, we extract $31\times 31$ patches from the boundary areas of the reference section and its corresponding contrast map and group similar patches into texture tensors using the proposed classification method in Algorithm~\ref{algo:algo1}. The dimension of the subspace is $[15,15,5]$, and threshold $T_e$ for the reconstruction error in the classification of texture tensors is $3$. Furthermore, in the predicted sections, we search along the normal directions of projected points and localize tracked points by comparing the similarity between the patches of candidate points and texture tensors built from the reference section. The searching process and the connection of tracked points are implemented automatically, which improves interpretation efficiency.

\begin{figure}[t]
\centering
\begin{minipage}[b]{1.0\linewidth}
\centering
\centerline{\includegraphics[height=3cm]{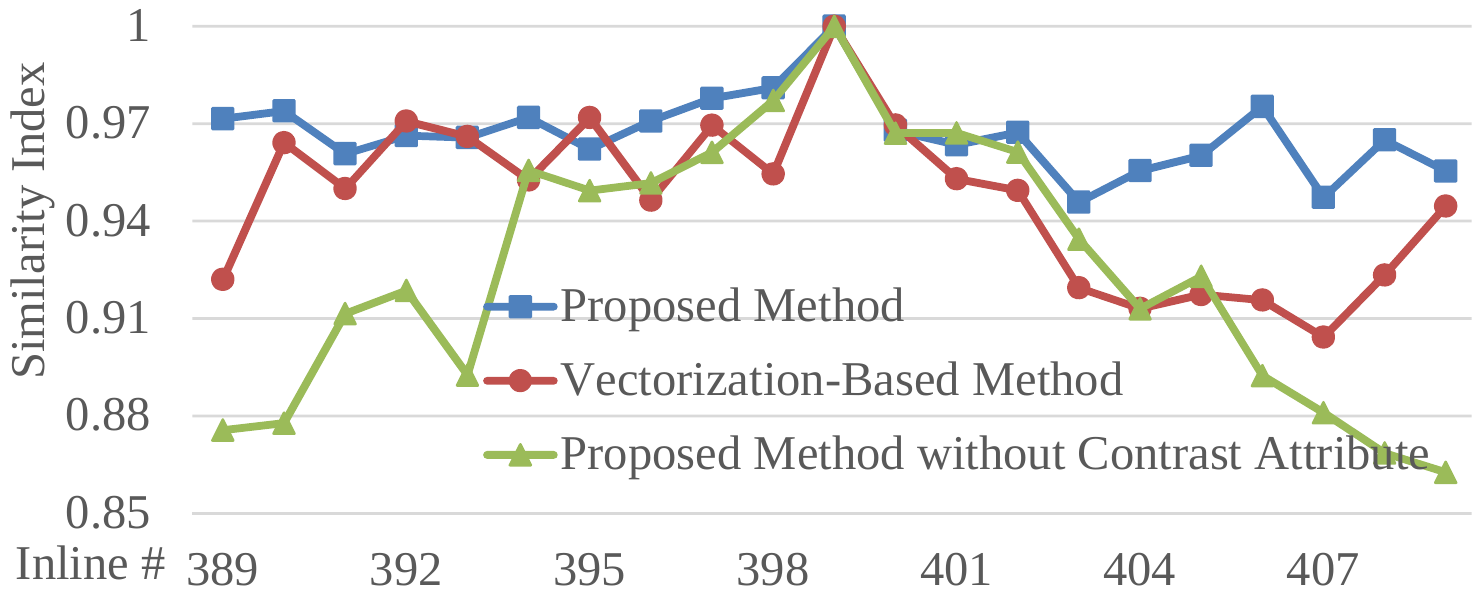}}
\end{minipage}
\caption{The comparison of tracked boundaries synthesized by different methods.}
\label{fig:cmpr1}
\end{figure}

\begin{figure}[t]
\centering
\begin{minipage}[b]{1.0\linewidth}
\centering
\centerline{\includegraphics[height=3cm]{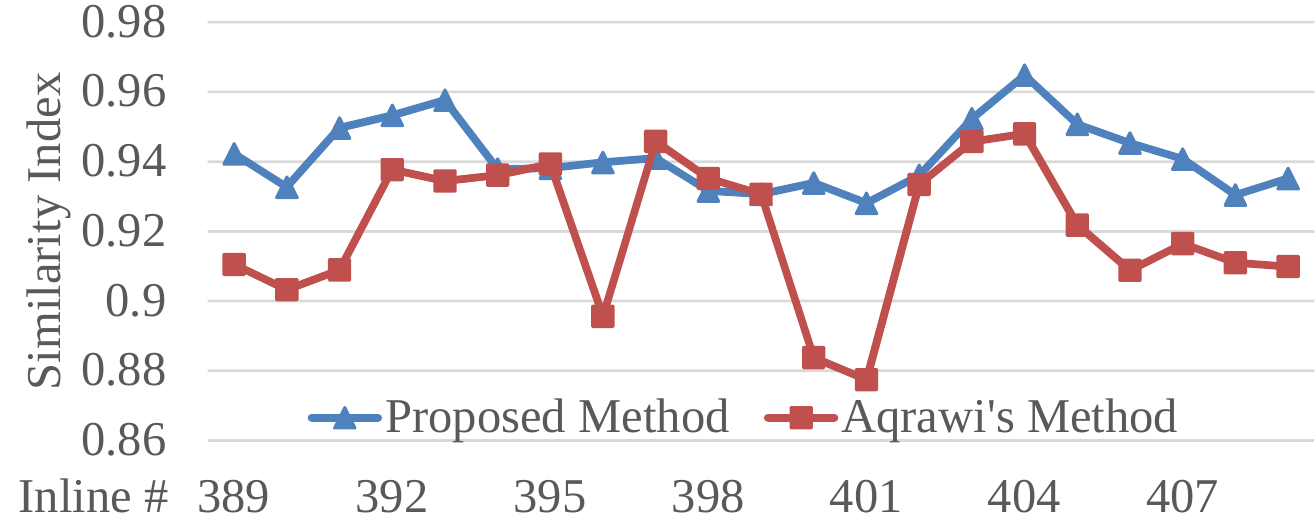}}
\end{minipage}
\caption{The comparison of boundaries synthesized by the proposed method and detected by Aqrawi's method.}
\label{fig:cmpr2}
\end{figure}

\begin{figure}[t]
\begin{minipage}[b]{0.49\linewidth}
  \centering
  \centerline{\includegraphics[height=1.7cm]{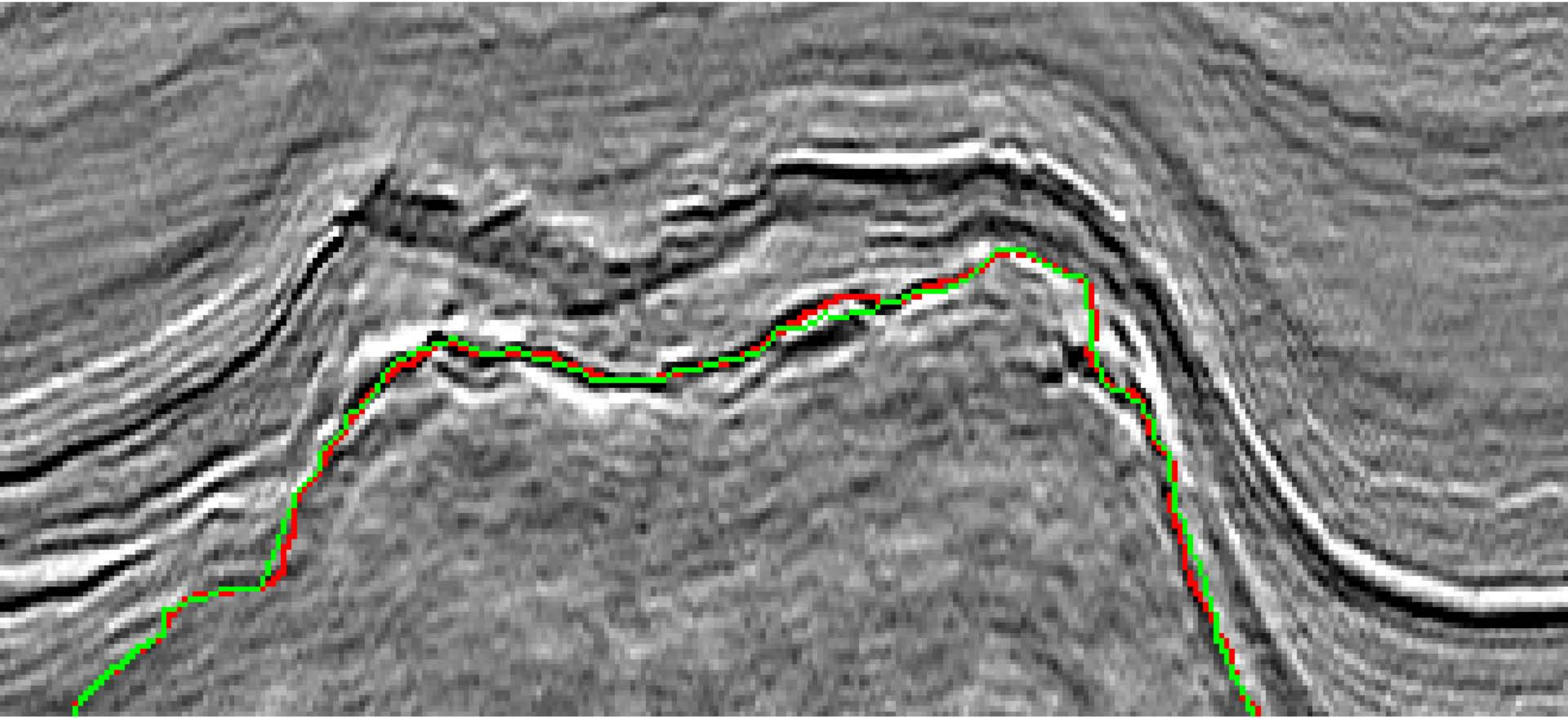}}
  \centerline{(a)}\medskip
\end{minipage}
\begin{minipage}[b]{0.24\linewidth}
  \centering
  \centerline{\includegraphics[height=1.7cm]{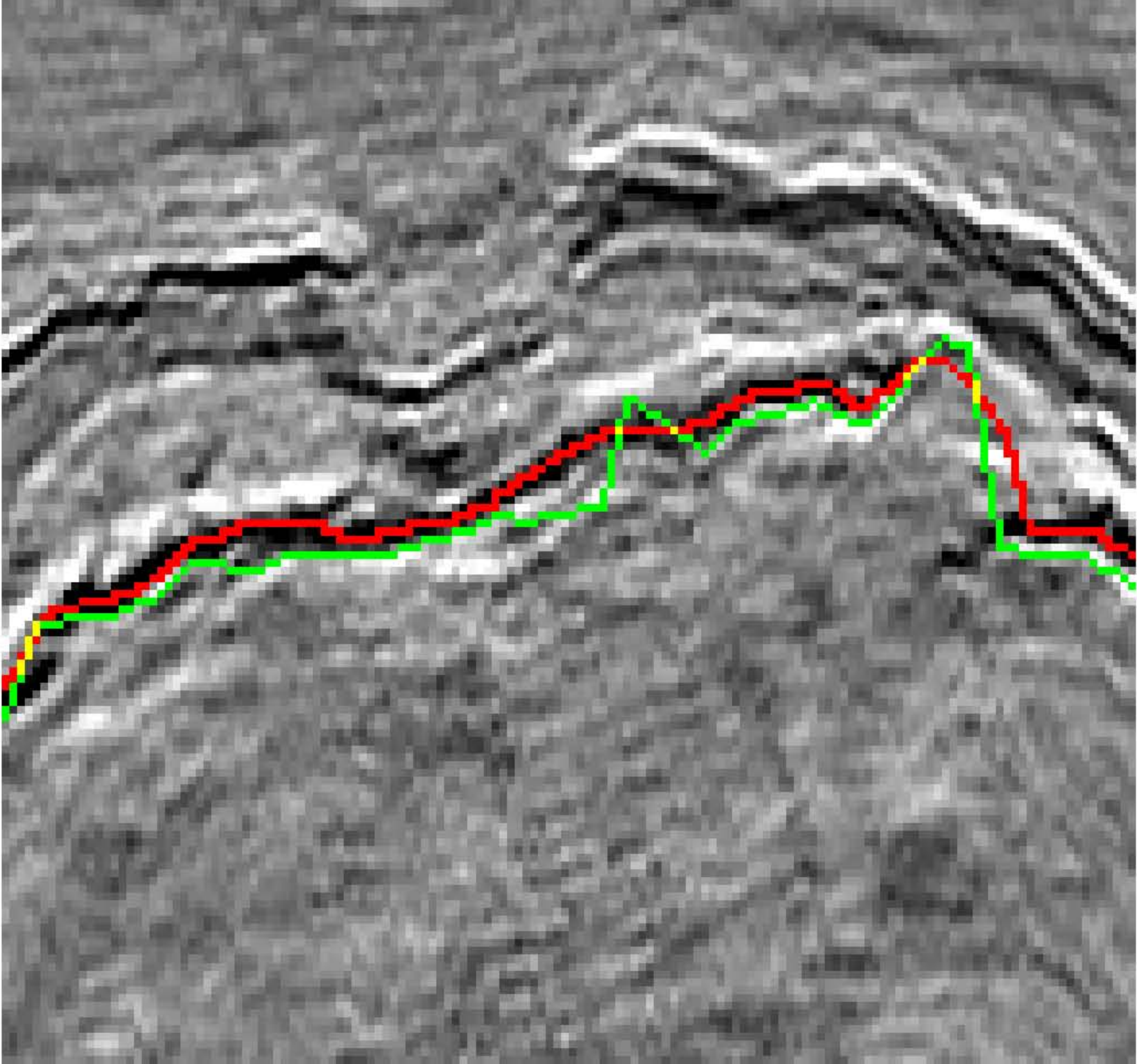}}
  \centerline{(b)}\medskip
\end{minipage}
\begin{minipage}[b]{0.24\linewidth}
  \centering
  \centerline{\includegraphics[height=1.7cm]{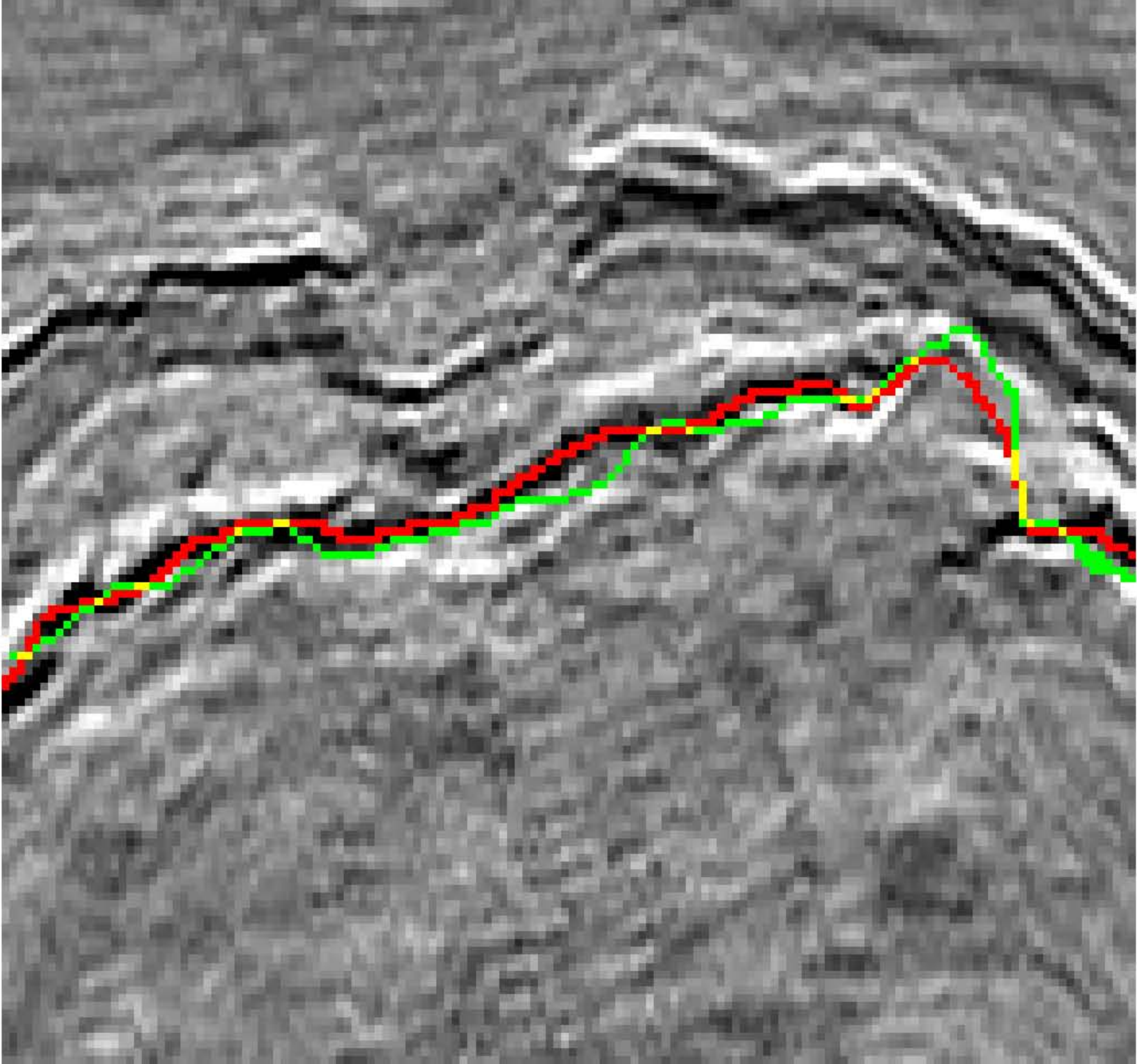}}
  \centerline{(c)}\medskip
\end{minipage}
\caption{(a) The tracked boundary (green) and the ground truth (red) in Inline \#409, (b) The boundary (green) detected by Aqrawi's method and the ground truth (red) in Inline \#392, and (c) The boundary (green) synthesized by the proposed method and the ground truth (red) in Inline \#392.}
\label{fig:cmprs}
\end{figure}

To evaluate the necessity of the contrast attribute and the tensor-based learning process in salt-dome tracking, we track the manually labeled boundary through predicted sections using three methods, the proposed method, the proposed method without the contrast attribute, and the vector-based subspace learning method. The last one refers to a method that extracts features only from the subspace of vectorized texture patches, which constitute the 3-mode unfolding matrices of tensors in the proposed method. To objectively evaluate the similarity between tracked boundaries and the ground truth, we calculate Fr\'echet distances~\cite{alt1995computing} between the segments of curves and normalize the averaged distance as a similarity index. Fig.~\ref{fig:cmpr1} illustrates the similarity indices of salt-dome boundaries labeled by the three methods. The proposed method synthesizes tracked boundaries with the highest accuracy, particularly in predicted sections with larger inline offsets. For example, in Fig.~\ref{fig:cmprs}(a), the green tracked boundary in Inline \#409 almost coincides with the ground truth, labeled by the red curve. In addition, to verify the robustness of the proposed method in practical cases, we detect the salt-dome boundary of the reference section by Aqrawi's method~\cite{aqrawi2011detecting} and synthesize tracked boundaries in predicted sections. In Fig.~\ref{fig:cmpr2}, tracked boundaries in most of the predicted sections have higher similarity indices than those detected by Aqrawi's method, which indicates the higher reliability of the proposed method. In Figs.~\ref{fig:cmprs}(b) and (c), we compare tracked and detected boundaries (green) with the ground truth (red), respectively, and find that the former has more deviations.


\vspace{-0.1in}
\section{Conclusion}
\label{sec:conclude}
\vspace{-0.05in}
In this paper, we developed the tracking of salt-dome boundaries using a tensor-based subspace learning algorithm. We built texture tensors by classifying image patches along the boundary areas of the reference section and its corresponding contrast map. With features extracted from the subspace of texture tensors, we identified the positions of tracked points by evaluating their similarity to texture tensors. The connection of tracked points synthesizes tracked boundaries. Experimental results showed that to obtain more accurate labeling, the tracking method needs to employ contrast maps and tensor-based analysis. In practical cases, the proposed method also showed higher reliability than the state-of-the-art detection method.

\vspace{-0.1in}
\section{Acknowledgements}
\label{sec:ack}
This work is  supported by the Center for Energy and Geo Processing (CeGP) at Georgia Tech and by King Fahd University of Petroleum and Minerals (KFUPM).

\bibliographystyle{IEEEbib}
\bibliography{main}

\end{document}